\ifx\onecol\undefined
\documentclass[journal,twocolumn]{IEEEtran}
\else 
\documentclass[12pt,draftclsnofoot,journal,onecolumn]{IEEEtran}
\fi
\usepackage{amsfonts}
\usepackage{amsmath,amssymb}
\usepackage{acronym}  
\usepackage{algorithm}
\usepackage{algorithmic}
\usepackage{bm}
\usepackage{bbm}
\usepackage{booktabs}
\usepackage{color, soul}
\usepackage{cite}
\usepackage{graphicx}
\usepackage{indentfirst}
\usepackage{lipsum}
\usepackage{setspace}
\usepackage{tikz}
\usetikzlibrary{arrows}
\usepackage{subfigure}
\usepackage[amsmath,thmmarks]{ntheorem}
\usepackage{theorem}
\usepackage{url}

\begin{document}

\title{Large Language Model Enabled Multi-Task Physical Layer Network}

\author{Tianyue~Zheng, {\textit{Graduate Student Member, IEEE}}, and Linglong~Dai, {\textit{Fellow, IEEE}}
\thanks{
This work was supported in part by the National Science Fund for Distinguished Young Scholars (Grant No. 62325106), in part by the Key Program of the National Natural Science Foundation of China (Grant No. 62031019), and in part by the National Key Research and Development Program of China (Grant No. 2023YFB3811503). 
}
\thanks{T. Zheng, and L. Dai are with the Department of Electronic Engineering, Tsinghua University, and the State Key laboratory of Space Network and Communications, Tsinghua University, Beijing 100084, China (e-mails: zhengty22@mails.tsinghua.edu.cn, daill@tsinghua.edu.cn). L. Dai is also with Department of Electrical Engineering and Computer Science, Massachusetts Institute of Technology, Cambridge, MA 02139, United States (Email: daill@mit.edu).
}}

\maketitle
\begin{abstract}
The advance of Artificial Intelligence (AI) is continuously reshaping the future 6G wireless communications. 
Particularly, the development of Large Language Models (LLMs) offers a promising approach to effectively improve the performance and generalization of AI in different physical-layer (PHY) tasks.
However, most existing works finetune dedicated LLM networks for a single wireless communication task separately. Thus performing diverse PHY tasks requires extremely high training resources, memory usage, and deployment costs.
To solve the problem, we propose a LLM-enabled multi-task PHY network to unify multiple tasks with a single LLM, by exploiting the excellent semantic understanding and generation capabilities of LLMs.
Specifically, we first propose a multi-task LLM framework, which finetunes LLM to perform multi-user precoding, signal detection and channel prediction simultaneously. 
Besides, multi-task instruction module, input encoders, as well as output decoders, are elaborately designed to distinguish different tasks. The proposed design allows different wireless data types to be well aligned with the LLM input format. Moreover, low-rank adaptation (LoRA) is utilized for LLM fine-tuning. To reduce the memory requirement during LLM fine-tuning, a LoRA fine-tuning-aware quantization method is introduced.
Extensive numerical simulations are also displayed to verify the effectiveness of the proposed method.
\end{abstract}

\begin{IEEEkeywords}
    large language models (LLMs), multi-task LLM, physical layer communications
\end{IEEEkeywords}

\vspace{-1em}
\section{Introduction} \label{sec-intro}
The deep integration of Artificial Intelligence (AI) with wireless communications is one of the key features of Sixth-Generation (6G) communications~\cite{AI6G}.
With the development of 6G communications, the application of AI becomes essential to manage the explosively-growing data service.
Attributed to the strong feature extraction capability, AI, especially deep learning (DL), has demonstrated great potential in a wide range of physical-layer (PHY) communication tasks~\cite{DL}, including CSI feedback~\cite{CSIFEED}, channel estimation~\cite{CE}, channel prediction~\cite{Informer}, signal detection~\cite{OAMP}, etc.
The AI Radio Access Network (AI-RAN) is anticipated to offer reduced latency, improved bandwidth, spectrum efficiency, and coverage. 
Moreover, significant efforts have been made by 3rd Generation Partnership Project (3GPP) to standardize DL in wireless networks recently~\cite{stand}.

Despite the significant progress, existing DL empowered methods still face some fundamental issues that limit their applications in practical communication networks. 
First, with the dramatically increasing channel dimension and rapidly changing wireless environment in 6G applications, it is difficult for existing DL methods to comprehensively recognize patterns from complicated data distribution, due to their small model size and simple network structure.
This insufficiency restricts their ability to provide reliable solutions in dynamic real-world scenarios.
Secondly, existing DL methods exhibit a poor generalization capability to different wireless environments. For instance, a channel prediction model trained in an indoor scenario usually requires retraining when the wireless environment is changed to urban.
Fortunately, recent advancements in large language models (LLMs)~\cite{gpt,Llama2,gpt3} have provided a radical solution to the challenges of existing DL methods.
Bearing a huge amount of parameters, LLMs possess the ability to capture universal knowledge, which has demonstrated impressive language understanding and generation capabilities to various tasks in different domains.

Very recently, several initial studies~\cite{power,LLMCSIFeed,LLM4CP,BP} have been conducted since 2024 to leverage LLMs for boosting the performance of PHY communication systems. To be specific, in~\cite{power} LLM is adopted to perform power allocation using a few-shot learning approach. In this method, the channel gains and the corresponding transmit power strategies are provided as the input of LLM, also called  ``prompt" in the society of AI. The paper~\cite{power} demonstrates that LLM can automatically understand the principle of water-filling based optimal power allocation without any retraining. 
The authors in~\cite{LLMCSIFeed} effectively utilize LLMs to enhance AI-based CSI feedback performance in various scenarios. They incorporate the channel distribution as a prompt within the decoder to further enhance channel reconstruction quality.
Besides, in order to obtain more accurate channel prediction and improve the generalization capability, the authors in~\cite{LLM4CP} propose a LLM-driven channel prediction approach.
~\cite{BP} also unleashes the strength of LLMs for time series forecasting to improve the robustness of beam prediction. 
Moreover, a few review papers~\cite{LLM1,LLM2,LLM3,LLM4,LLM5} have explored the potential transformative impact of LLMs and provide envisions for LLM-aided wireless communications.

However, most existing works finetune dedicated LLM networks for a {\it single} wireless communication task such as the previously mentioned channel prediction and CSI feedback, etc. In reality, wireless communication systems involve a multitude of tasks, each with distinct requirements that necessitate the adoption of LLMs. 
Given the huge model size of LLM, designing and training dedicated models for each task separately will lead to extremely high computational complexity, memory usage, and deployment costs.

To solve the problem, we propose a LLM-enabled multi-task PHY network to {\it unify} multiple tasks with a {\it single} LLM, by leveraging the excellent language understanding and generation capabilities of LLMs.
The main contributions of this paper are summarized as follows\footnote{Simulation codes will be provided to reproduce the results in this paper: \url{http://oa.ee.tsinghua.edu.cn/dailinglong/publications/publications.html}.}.
\begin{itemize}
    \item We propose a multi-task LLM framework for PHY communications. The framework enables us to input different task requirements to LLMs with natural language, and explore the global feature extraction ability of LLMs to execute multiple tasks within one network.
    Particularly, this work focuses on three of the most typical tasks in PHY communications, multi-user precoding, signal detection, and channel prediction. The proposed framework can also accommodate any other PHY tasks. 
    \item For the design of the proposed framework, dedicated modules are proposed to adapt the LLMs for multiple tasks in the wireless domain. It mainly consists of two components: multi-task instructions, and task-specific encoders and decoders. Specifically, first, in order to distinguish and cope with different tasks and different data formats of the tasks, we elaborately design multi-task instructions as prompts of LLM; secondly, to bridge the gap between the feature of wireless data and that of the LLM, task-specific encoders and decoders are proposed to adapt the text-based pre-trained LLM.
    \item For the fine-tuning of the proposed framework, we introduce low-rank adaptation (LoRA)~\cite{lora}, which freezes the pre-trained model weights and injects trainable low-rank matrices. Furthermore, to mitigate the computational and memory demands of the proposed model, we employ a LoRA fine-tuning-aware quantization method~\cite{loftq} that simultaneously quantizes an LLM and finds a proper low-rank initialization for LoRA fine-tuning.
    \item Extensive simulation experiments have been conducted to verify the effectiveness of the proposed method, which achieves comparable performance with the dedicated designed network for each task. In addition, the LoRA fine-tuning-aware quantization method achieves 75\% storage reduction during fine-tuning with almost no performance degradation. Moreover the proposed multi-task instruction effectively accelerates the convergence and improve the performance.
\end{itemize}

The rest of the paper is organized as follows. Section II introduces the system model. Then the three selected tasks, namely multi-user precoding, signal detection, and channel prediction, are formulated respectively. In Section III, the design of the proposed multi-task PHY LLM is illustrated in detail. 
Besides, Section IV elaborates the fine-tuning strategy of the proposed multi-task PHY LLM.
Simulation results are provided in Section V. Finally, Section VI concludes this paper.

{\it Notation:} ${\bf a}^H$, ${\bf A}^H$ denote the conjugate transpose of vector $\bf a$ and matrix $\bf A$, respectively; $\left\| {\bf a} \right\|_2$ denotes the $l_2$ norm of vector $\bf a$; $\left\| {\bf A} \right\|_F$ denotes the Frobenius norm of matrix $\bf A$; $\mathbb{R}$,$\mathbb{C}$ denote the set of real numbers and complex numbers, respectively; $\mathcal{C} \mathcal{N} ({\bf \mu},{\bf \Sigma})$ denotes the probability density function of complex multivariate Gaussian distribution with mean ${\bf \mu}$ and variance ${\bf \Sigma}$.

\section{Systems Model}\label{sec-sys}
We consider a multi-user (MU) multiple-input-single-output (MISO)- orthogonal-frequency-division-multiplexing (OFDM) system working in a time-division-duplex (TDD) mode. A base station (BS) simultaneously serves $K$ single-antenna mobile users using $K$ RF chains. The BS is equipped with a uniform planar array (UPA)  consisting of $N_{\rm T}=N_h \times N_v$ antennas, where $N_h$ and $N_v$ denote the quantity of antennas along the horizontal and vertical dimensions, respectively. To design a multi-task LLM for the BS, we select three typical tasks in PHY communications for instance, downlink multi-user precoding, uplink signal detection, and channel prediction for mobile devices, respectively. The system models and problem descriptions of these three tasks are presented one by one below. 

\subsection{Multi-user Precoding}
For downlink transmission scenario, the channel between user $k$ and the BS at the $m$-th subcarrier is denoted as ${{\bf h}_k^m} \in \mathbb{C}^{N_t \times 1}, m=1,2,\cdots,M$. The received signal of user $k$ at the $m$-th subcarrier is given by
\begin{equation}
    y_{k}^m = {\mathbf{h}_{k}^{m}}^{H} \sum_{k'=1}^{K} \mathbf{w}_{k'}^m x_{k'}^m + n_{k}^m,
\end{equation}
where $\mathbf{w}_{k}^m$ represents the beamforming vector for user $k$, $x_{k}^m$ with $\mathbb{E}(|x_k^m|^2) = 1$, is the transmitted symbol from the BS to user $k$, and $n_{k}^m \sim \mathcal{CN}(0,\sigma^2)$ denotes the additive Gaussian white noise (AWGN) with zero mean and variance $\sigma^2$. 

Multi-user precoding aims to maximize the system sum rate via the optimization of the transmit precoders.  The total power of all beamforming vectors is limited due to the BS power budget.
For simplicity, we design the beamformers based on the channel of the central carrier-frequency  ${\bf h}_k$ and the problem is mathematically formulated as
\begin{align}\label{prob-pre}
\max_{\mathbf{W}} \sum_{k=1}^{K} \log_2(1 + \gamma_k), \quad
 \text{s.t. } \sum_{k=1}^{K} \|\mathbf{w}_k\|^2 \leq P_{\max},
\end{align}
where $\mathbf{W} = [\mathbf{w}_{1},\mathbf{w}_{2},\cdots,\mathbf{w}_{K}]$ is a set of beamforming vectors and $P_{\rm max}$ is the power budget. 
Besides, $\gamma_k$ represents the received signal-to-interference-plus-noise ratio (SINR) at user $k$. It is written as 
\begin{equation}
    \gamma_{k} = \frac{\left| \mathbf{h}_{k}^{H} \mathbf{w}_{k} \right|^2}{\sum_{k'=1, k' \neq k}^{K} \left| \mathbf{h}_{k}^{H} \mathbf{w}_{k'} \right|^2 + \sigma^2}.
\end{equation}

As pointed out in~\cite{prestruc}, the optimal downlink beamforming vectors for~\eqref{prob-pre} follow the structure as
\begin{equation}\label{eq-pre}
    \mathbf{w}_k^* = \sqrt{p_k} \frac{\left( \mathbf{I}_{N_{\rm T}} + \sum_{k=1}^{K} \frac{\lambda_k}{\sigma^2} \mathbf{h}_k \mathbf{h}_k^H \right)^{-1} \mathbf{h}_k}{\left\| \left( \mathbf{I}_{N_{\rm T}} + \sum_{k=1}^{K} \frac{\lambda_k}{\sigma^2} \mathbf{h}_k \mathbf{h}_k^H \right)^{-1} \mathbf{h}_k \right\|_2}, \quad \forall k,
\end{equation}
where $p_k$ is the power allocated to the $k$-th use, and ${\lambda_k}$ is a positive parameter and $\sum_{k=1}^{K} {\lambda_k} = \sum_{k=1}^{K} p_k = P_{\rm max}$. The solution structure in~\eqref{eq-pre} provides the required expert knowledge for the LLM-empowered beamforming design in~\eqref{prob-pre}. Given this knowledge, the LLM is only required to learn $2K$ key parameters $\mathbf{\lambda} = [\lambda_1,\lambda_2,\cdots,\lambda_K]$ and $\mathbf{p} = [p_1,p_2,\cdots,p_K]$, instead of the whole $K\times N$ beamforming matrix $\mathbf{W} = [\mathbf{w}_{1},\mathbf{w}_{2},\cdots,\mathbf{w}_{K}]$.
Thus, the problem of DL-based multi-user precoding can be reformulated as
\begin{align}
\textbf{P1: } \min_{\Omega_{\rm PRE}} \quad & \sum_{k=1}^{K} \log_2(1 + \gamma_k),\\
\text{s.t.} \quad & \mathbf{w}_k ~{\rm in~\eqref{eq-pre}},~ \hat{{\mathbf{\lambda}}},\hat{{\mathbf{p}}} = f_{\Omega_{\rm PRE}}(\mathbf{H}), 
\end{align}
where $\mathbf{H}=[\mathbf{h}_1,\mathbf{h}_2,\cdots,\mathbf{h}_K]$, and $f_{\Omega_{\rm PRE}}$ is the mapping function with variable parameters $\Omega_{\rm PRE}$.

\subsection{Signal Detection}
During uplink transmission, the BS antenna array simultaneously receive the transmitted symbols from the $K$ users. 
For clarity, we introduce the superscript ``{$\bar{\cdot}$}" to all parameters related to the uplink transmission, e.g., $\bar{a}$. 
Specifically, we denote the transmitted symbol vector from all users at the $m$-th subcarrier as $\bar{\mathbf{x}}^m = [\bar{x}^m_1, \bar{x}^m_2, \cdots, \bar{x}^m_K] \in \mathbb{C}^{K \times 1}$.
Each element is drawn from the $P$-QAM constellation, and then  transmitted over channel. The received signal $\bar{\mathbf{y}}^m \in \mathbb{C}^{N_{\rm T} \times 1}$ is given by
\begin{equation}
    \bar{\mathbf{y}}^m = \bar{\mathbf{H}}^m \bar{\mathbf{x}}^m + \bar{\mathbf{n}}^m,
\end{equation}
where $\bar{\mathbf{H}}^m = [\bar{\mathbf{h}}_1^m,\bar{\mathbf{h}}_2^m.\cdots,\bar{\mathbf{h}}_K^m]$ is the uplink channel of the $K$ users and $\bar{\mathbf{n}}^m \sim \mathcal{CN}(0,\bar{\sigma}^2 \mathbf{I}_{N_{\rm T}})$ is the AWGN at the $m$-th subcarrier.

BS requires to recover the signals $\bar{\mathbf{x}}^m$ from the received signal $\bar{\mathbf{y}}^m$ given $\bar{\mathbf{H}}^m$. We adopt the minimum mean squared error (MMSE) estimator to formulate the associated signal detection problem as
\begin{align}
\textbf{P2: } \min_{\Omega_{\rm DET}} \quad & ||\hat{\bar{\mathbf{x}}}^m - \bar{\mathbf{x}}^m||_2 \\
\text{s.t.} \quad & \hat{\bar{\mathbf{x}}}^m = f_{\Omega_{\rm DET}}(\bar{\mathbf{H}}^m,\bar{\mathbf{y}}^m),
\end{align}
where $f_{\Omega_{\rm DET}}$ is the mapping function with variable parameters $\Omega_{\rm DET}$.

\subsection{Channel Prediction}
In mobile scenarios involving high-velocity users, it is possible that the channel coherence time is shorter than the channel estimation period. 
Under these circumstances, precise channel prediction becomes crucial to mitigate the channel aging phenomenon~\cite{aging} in high-mobility communication environments with rapidly changing channels.
In this work, we aim to accurately predict channel state information (CSI) of next $T_2$ time slots given CSI of previous $T_1$ time slots. The CSI of $M$ subcarriers at time $t$ is represented in matrix form as
\begin{equation}
    \mathbf{H}_k^t = [\mathbf{h}_k^{1,t},\mathbf{h}_k^{2,t},\cdots,\mathbf{h}_k^{M,t}], \forall k,t,
\end{equation}
where $\mathbf{h}_k^{m,t}$ is the channel of user $k$, time $t$, and subcarrier $m$.

To evaluate the channel prediction accuracy, the normalized MSE (NMSE) between predicted CSI by the network and ground-truth CSI is selected as the metric.
Utilizing the metric, the channel prediction problem can be described as follows:
\begin{align}
\textbf{P3: } \min_{\Omega_{\rm CP}} \quad &  \frac{\sum_{t=1}^{T_2} \| \hat{\mathbf{H}}^{t_0+t} - \mathbf{H}^{t_0+t} \|_F^2}{\sum_{t=1}^{T_2} \| \mathbf{H}^{t_0+t} \|_F^2}  \\
\text{s.t.} \quad &  (\hat{\mathbf{H}}^{t_0+1}, ..., \hat{\mathbf{H}}^{t_0+T_2}) = f_{\Omega_{\rm CP}}(\mathbf{H}^{t_0}, ..., \mathbf{H}^{t_0-T_1+1}),
\end{align}
where $\hat{\mathbf{H}}^{t}$ is the predicted CSI, and $f_{\Omega_{\rm CP}}$ is the mapping function of the network with trainable parameters ${\Omega_{\rm CP}}$. Note here we ignore the subscript since the channel prediction is performed independently to each user. 

\section{The Framework of the Proposed Physical Layer Multi-task LLM}\label{sec-pro}
\begin{figure}
	\centering 
	\includegraphics[width= \linewidth]{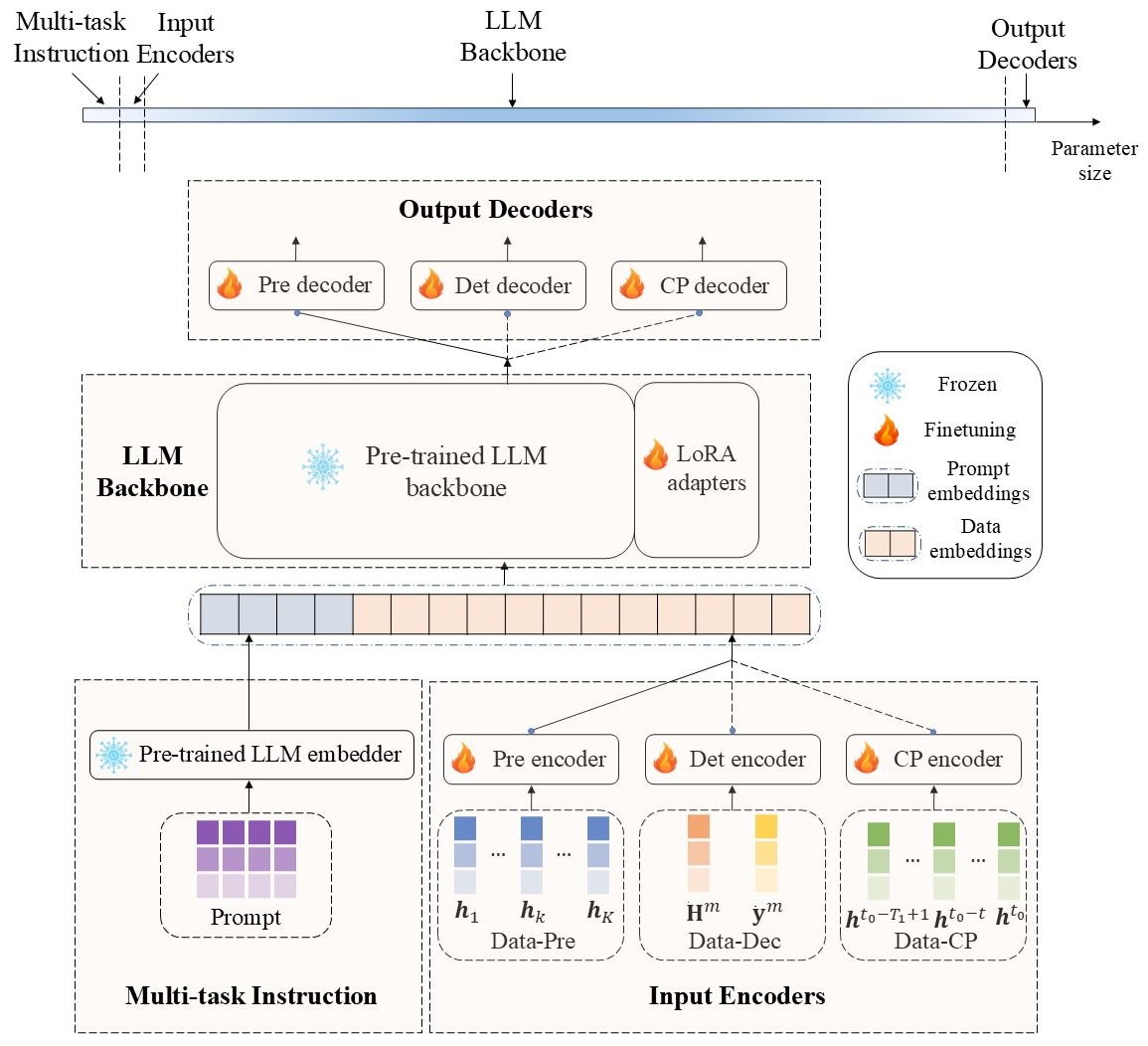}
        \caption{The model framework of our method. In this figure, ``CP" denotes channel prediction, ``Det" denotes signal detection, and ``Pre" denotes multi-user precoding.}
	\label{pic_frame}
\end{figure}

In this paper, we propose an LLM-enabled PHY network to {\it unify multiple tasks} (channel prediction, multi-user precoding, as well as signal detection) with a {\it single} network. The proposed framework is illustrated in Fig.~\ref{pic_frame}, including a multi-task instruction module, an input encoder, an LLM backbone, and an output decoder. 
To distinguish and cope with different tasks, we design multi-task instructions as prompts of LLM, which are processed by the pre-trained LLM embedder as part of the LLM input.
Besides, the wireless data are encoded by task-specific encoders to make the feature of \emph{wireless data} understandable to a \emph{text-based} pre-trained LLM, e.g., GPT-2. The prompt and encoded wireless data are concatenated together to serve as the inputs of LLM. These inputs are later fed into the LLM backbone. Note that the same LLM backbone is used for all considered tasks. 
Finally, obtaining the outputs of the LLM backbone, task-specific decoders are utilized to generate desired outputs for different tasks.

We describe each component of the framework as follows. 
It is worth noting that, in this paper, we select three typical tasks suitable for deep learning methods in PHY communications, namely multi-user precoding, signal detection, and channel prediction; while the proposed method can be smoothly extended to other PHY tasks.
Besides, we mention that although the encoders and decoders are designed specifically for each task, the parameter sizes of these modules are much smaller than that of the LLM backbone. The LLM backbone, which takes up the majority of the total parameter size, is shared among different tasks.

\subsection{Multi-task Instruction Template}\label{sec3-a}
When training a single unified model for multiple different tasks, a basic and critical problem is how to distinguish each task. Thanks to the remarkable text understanding capability, LLM allows users to input task requirements with natural language. Therefore, we can design a multi-task instruction template with task-specific tokens as prompt to make each task easily distinguishable. Next, we introduce our multi-task instruction template in detail.

We structure the instruction template into three parts, which is denoted as follows, 
\begin{equation}
    [Task~Identifier] {\rm Task~description} <Instruction>.
\end{equation}
The first part is the task identifier token, the second part is the task and data description, and the third part is the instruction input. \textbf{Task Identifier} provides a distinct identifier for each task to reduce the ambiguity across various tasks. \textbf{Task description} attempts to input basic domain knowledge and illustrate the dataset for LLM to better understand the task. Take multi-user precoding as an example, the designed task description is presented by ``{\it Multi-user precoding aims to maximize the sum rate of multiple users through the design of precoder. For the collected dataset, we consider a base station with 128 antennas to serve 8 single antenna users simultaneously}".
The third part \textbf{Instruction} gives direct and clear objective of the task. Again, for instance the instruction for multi-user precoding is ``{\it $<$Instruction$>$ Design the precoding matrix given channels of the users, to maximize the sum rate of the multiple users."}.

The designed prompt is then fed to the pre-trained LLM embedder as part of the input of LLM. The prompt embedding is denoted as $\mathbf{X}_{\rm promot}^{\rm emb}$.

\subsection{Input Encoders}
Adapting text-based pre-trained LLM to multiple communication domain tasks is another challenging problem. To be specific, first,
there is a huge gap between the specific characteristics of the data in the communication domain and the natural language. Therefore, directly applying a general-domain LLM to these communication tasks may lead to poor performance. Secondly, the format, length, and meaning of the data also differs significantly among the tasks.
The difficulties urge us to design dedicated input encoder modules for each task to fit the feature space of the LLM, and facilitate multi-task feature extraction.

\begin{figure}
	\centering 
	\includegraphics[width= \linewidth]{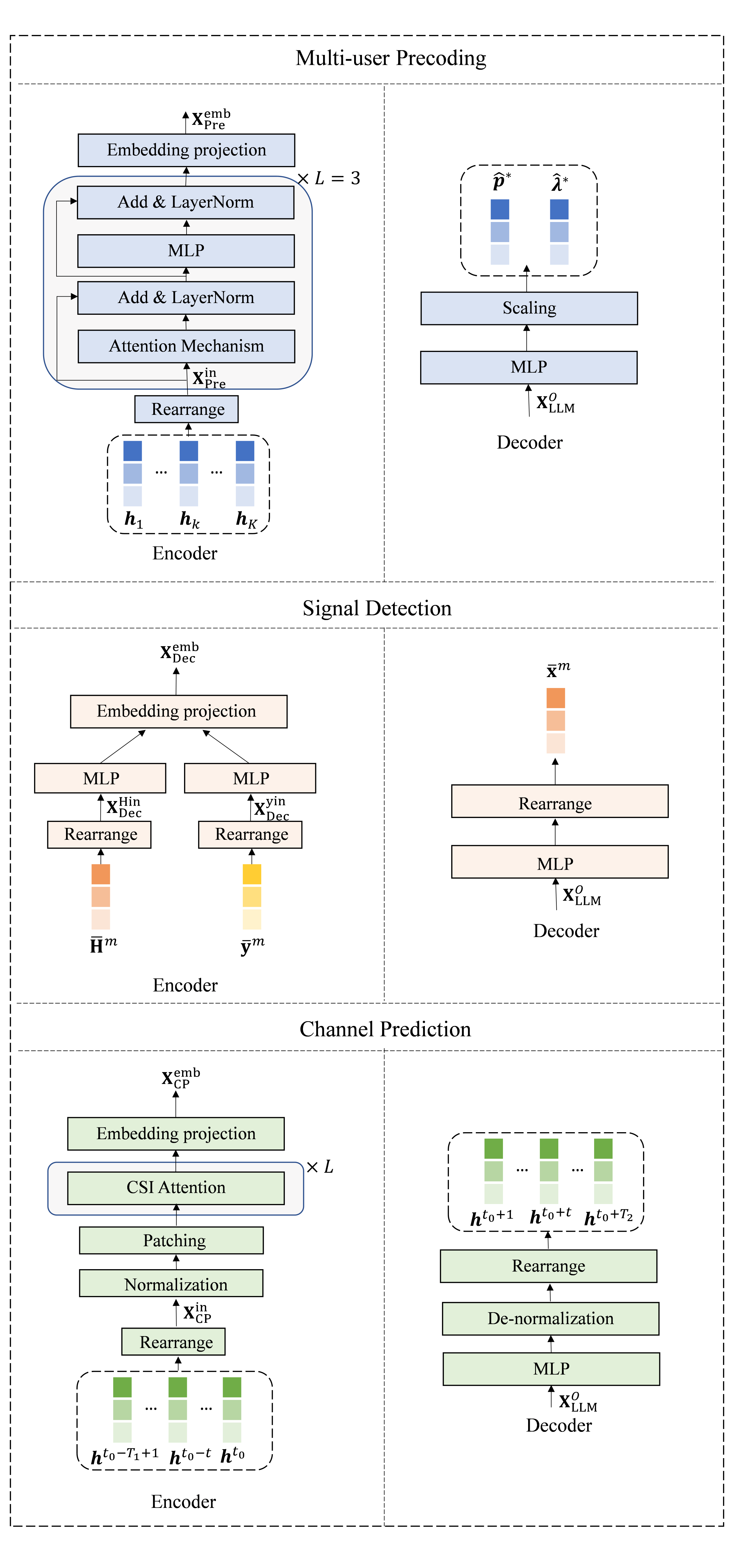}
        \caption{Detailed illustration of encoders and decoders.}
	\label{pic_ende}
\end{figure}

\subsubsection{Encoder for multi-user precoding} 
Since neural networks generally deal with real numbers, we first convert the complex multi-user channel into real tensors $\mathbf{X}_{\rm Pre}^{\rm in} \in \mathbb{R}^{K \times 2N_t}$ as the input.
For multi-user precoding, we employ a transfomer to extract the channel feature and inter-user relationship. The transfomer is composed of $L=3$ blocks. The structure of each block is presented in Fig.~\ref{pic_ende}, including a multi-head self-attention module and a multilayer perceptron (MLP) module.
The input $\mathbf{X}_{\rm Pre}^{\rm in}$ is first processed by the multi-head self-attention module:
\begin{equation}
    \mathbf{X}_{\rm Pre}^{\rm att(1)} = {\rm LayerNorm}({\rm ATT}(\mathbf{X}_{\rm Pre}^{\rm in}) + \mathbf{X}_{\rm Pre}^{\rm in}),
\end{equation}
where ${\rm ATT}(\cdot)$ denotes the multi-head self-attention.
Then MLP module is performed as
\begin{equation}
    \mathbf{X}_{\rm Pre}^{\rm mlp(1)} = {\rm LayerNorm}({\rm MLP}(\mathbf{X}_{\rm Pre}^{\rm att(1)})+\mathbf{X}_{\rm Pre}^{\rm att(1)}),
\end{equation}
which serves as the input of next block.
Denote the output of the transfomer as $\mathbf{X}_{\rm Pre}^{\rm mlp(3)}$. Then a linear layer is used to project $\mathbf{X}_{\rm Pre}^{\rm mlp(3)}$ into the language model space
\begin{equation}
    \mathbf{X}_{\rm Pre}^{\rm emb} = {\rm Linear}(\mathbf{X}_{\rm Pre}^{\rm mlp(3)}).
\end{equation}

\subsubsection{Encoder for signal detection}
The input of signal-detection encoder contains the uplink channel $\bar{\mathbf{H}}^m$ and the received signal $\bar{\mathbf{y}}^m$. The encoder rearranges the channel $\bar{\mathbf{H}}^m$ as one token $\mathbf{X}_{\rm Det}^{\rm Hin} \in \mathbb{R}^{K \times 2 N_t}$. Then a MLP module is employed to extract features from the channel
\begin{equation}
    \mathbf{X}_{\rm Det}^{\rm Hmlp} = {\rm MLP}(\mathbf{X}_{\rm Det}^{\rm Hin}).
\end{equation}
We also extract shallow features from the rearranged real received signal $\mathbf{X}_{\rm Det}^{\rm yin}$ by a MLP module
\begin{equation}
    \mathbf{X}_{\rm Det}^{\rm ymlp} = {\rm MLP}(\mathbf{X}_{\rm Det}^{\rm yin}).
\end{equation}
In the simulation parts, we find that integrating received signals of several time slots into one sample effectively improves the training performance compared to treating them as multiple samples. Therefore, we utilize received signals of $L_0=8$ time slots in one sample in practice. Then, the extracted features of channel and received data are concatenated and projected to the input format of LLM:
\begin{equation}
    \mathbf{X}_{\rm Det}^{\rm emb} = {\rm Linear}([ \mathbf{X}_{\rm Det}^{\rm Hmlp},\mathbf{X}_{\rm Det}^{\rm ymlp}]).
\end{equation}

\subsubsection{Encoder for channel prediction}
Consider that the CSI $\mathbf{H}^t \in \mathbb{C}^{N_t \times M}$ of at time $t$ is the high-dimensional structural data, directly predicting the matrix by the network will bring significant complexity. The complexity can be unacceptable for future 6G systems with a large number of antennas and subcarriers. Inspired by~\cite{LLM4CP}, we parallelize the channel prediction for different antennas.
That is to say, we predict the CSI of each transmitter antenna separately.  Thereby, the input sample of $j$-th antenna, for $j \in \{1,2,\cdots, N_T\}$, can be converted into a real tensor $\mathbf{X}_{\rm CP}^{\rm in} \in \mathbb{R}^{T_1 \times 2M}$. 
To facilitate convergence of network training, we first perform batch normalization for the input data as $\mathbf{X}_{\rm CP}^{\rm in'}$.
Then the normalized input $\mathbf{X}_{\rm CP}^{\rm in'}$ is divided into $T'_1 = \lceil\frac{T_1}{N} \rceil$ non-overlapping patches of size $N$ along the temporal dimension~\cite{patch}. The patching operation helps to capture the local temporal features and the patched input is denoted as $\mathbf{X}_{\rm CP}^{\rm pat} \in \mathbb{R}^{T'_1 \times N \times 2M}$.
We adopt the CSI attention module proposed in~\cite{SENET}, to extract preliminary temporal and frequency features before LLM:
\begin{equation}
    \mathbf{X}_{\rm CP}^{\rm CA} = {\rm CSIATT}^L (\mathbf{X}_{\rm CP}^{\rm pat}),
\end{equation}
where ${\rm CSIATT}^L$ represents the CSI attention module cascaded $L$ times. 
Finally, to align with the input format of LLM backbone, $\mathbf{X}_{\rm CP}^{\rm CA}$ is rearranged to $\mathbf{X}_{\rm CP}^{\rm CA'} \in \mathbb{R}^{T'_1 \times 2MN}$ and then mapped to  the feature dimension of the pre-trained LLM with a single fully-connected layer:
\begin{equation}
    \mathbf{X}_{\rm CP}^{\rm emb} = {\rm Linear}( \mathbf{X}_{\rm CP}^{\rm CA'}).
\end{equation}

\subsection{LLM Mainbody}\label{sec3-c}
\begin{figure}
	\centering 
	\includegraphics[width= \linewidth]{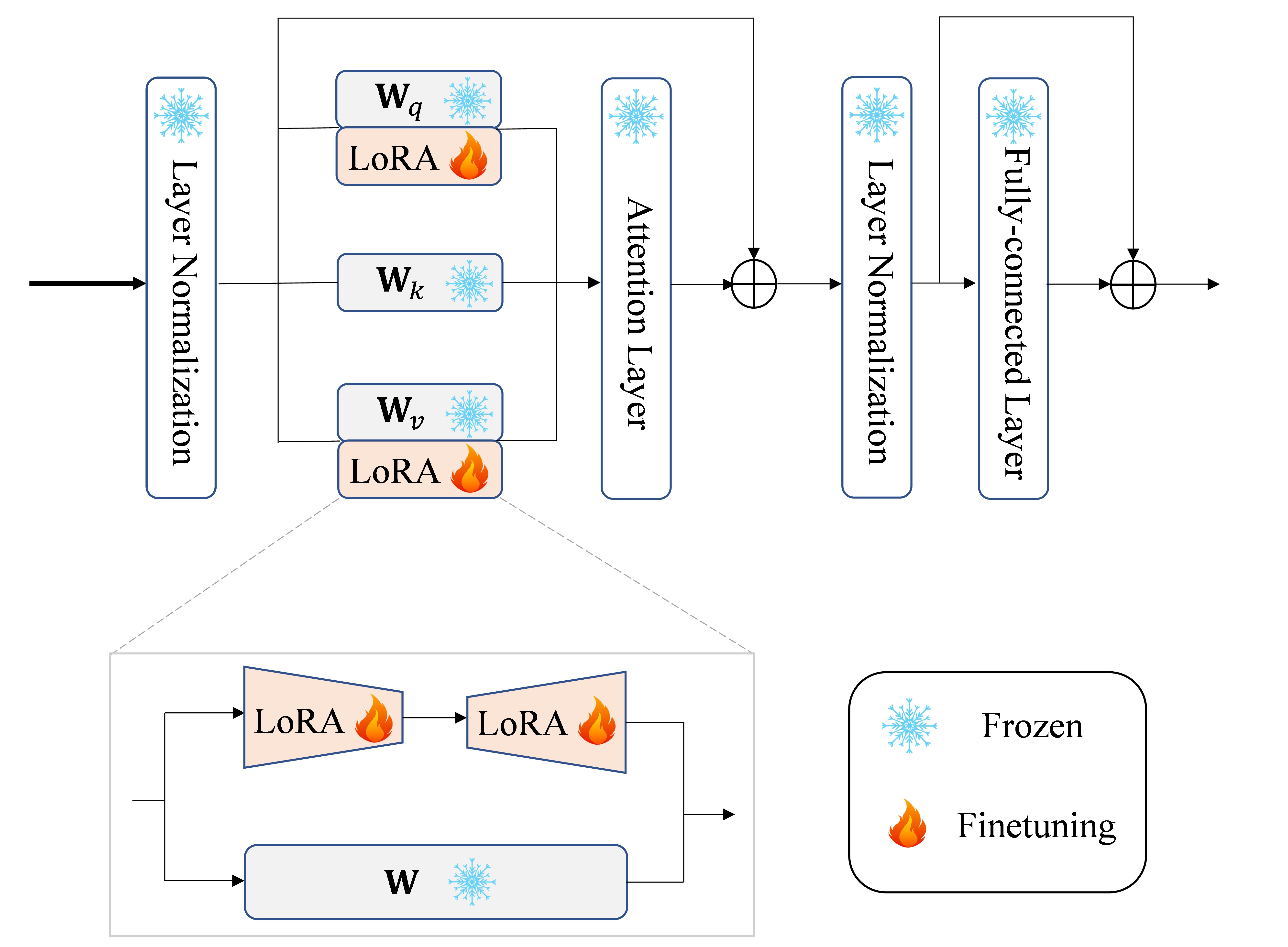}
        \caption{The illustration of a transformer block in LLAMA2.}
	\label{pic_lora}
\end{figure}

In this paper, we adopt LLAMA2 as the LLM backbone for our proposed multi-task PHY framework. It should be emphasized that our framework allows for seamless integration of alternative LLMs, including but not limited to GPT~\cite{gpt3} and QWen~\cite{qwen}. The decision of model architecture and scale requires evaluation of the trade-off between computational complexity and performance.

The architecture of LLAMA2 is presented in Fig.~\ref{pic_lora}. It consists of stacked classical  transformer decoders. 
Note that in the proposed multi-task PHY network, the LLM backbone takes up most of the model parameters. In contrast, the encoders and decoders, though designed specifically for each task, only occupies a very small portion in the network.
To be specific, the parameter size of LLAMA2-7B backbone is 7 billion, while the parameter size of all other modules, including the multi-task instructions, input encoders and output decoders, is only about 19 million.
Obtaining the LLM backbone, we concatenate the embedding of prompt and data obtained in the previous subsections as $\mathbf{X}^{\rm emb}$ and feed into the LLM backbone. The output can be obtained by
\begin{equation}
    \mathbf{X}^{O}_{\rm LLM} = {\rm LLM}(\mathbf{X}^{\rm emb}).
\end{equation}

\subsection{Output Decoders}
Similar to input encoders, to facilitate multiple downstream tasks simultaneously, task-specific decoders are required to convert the output features of the LLM into the final results for different tasks. Here, we elaborate on the output decoders of the three tasks sequentially. Note that for output decoders, we discard the prefix portion that is associated with the instruction prompt, while we only retain the output presentations of the data for the decoder.
\subsubsection{Decoder for multi-user precoding}
As discussed in Section~\ref{sec-sys} A., the multi-user precoding problem can be transformed into the learning of the parameters $\mathbf{\lambda} = [\lambda_1,\lambda_2,\cdots,\lambda_K]$ and $\mathbf{p} = [p_1,p_2,\cdots,p_K]$. Therefore, a MLP module comprising of two fully-connected (FC) layers is adopted to generates $2K$ values including the power allocation vectors $\hat{\mathbf{\lambda}}$ and $\hat{\mathbf{p}}$. The output of the MLP module $\mathbf{X}^{O}_{\rm Pre} \in \mathbb{R}^{K \times 2}$ can be written as
\begin{equation}
    \mathbf{X}^{O}_{\rm Pre} = {\rm MLP}(\mathbf{X}^{O}_{\rm LLM}),
\end{equation}
where $\hat{\mathbf{\lambda}} = \mathbf{X}^{O}_{\rm Pre}[:,1]$ and $\hat{\mathbf{p}} = \mathbf{X}^{O}_{\rm Pre}[:,2]$.  
Then the scaling layer scales the results of the output layer $\hat{\mathbf{\lambda}}$  and $\hat{\mathbf{p}}$ to meet the power constraint by:
\begin{equation}
    \hat{\mathbf{p}}^* = \frac{P_{\max}}{\|\hat{\mathbf{p}}\|_1} \hat{\mathbf{p}} \quad \text{and} \quad \hat{\mathbf{\lambda}}^* = \frac{P_{\max}}{\|\hat{\mathbf{\lambda}}\|_1} \hat{\mathbf{\lambda}}.
\end{equation}

\subsubsection{Decoder for signal detection}
For signal detection, we employ two FC layers to transform the dimension of the $\mathbf{X}^{O}_{\rm LLM}$ to the number of antennas:
\begin{equation}
    \mathbf{X}^{\rm mlp}_{\rm Det} = {\rm MLP}(\mathbf{X}^{O}_{\rm LLM}).
\end{equation}
Then $\mathbf{X}^{\rm mlp}_{\rm Det}$ is rearranged $\mathbf{X}^{O}_{\rm Det} \in \mathbb{R}^{K \times 2}$ to where the first and the second dimension respectively correspond to the real part and the imaginary part. 

\subsubsection{Decoder for channel prediction}
We utilize an MLP module as the decoder to predict the channel. The output is presented by
\begin{equation}
    \mathbf{X}^{\rm mlp}_{\rm CP} = {\rm MLP}(\mathbf{X}^{O}_{\rm LLM}).
\end{equation} 
Last, $\mathbf{X}^{\rm mlp}_{\rm CP}$ is de-normalized to generate the final output of the network, i.e.,
\begin{equation}
    \mathbf{X}^{norm}_{\rm CP} = \sigma_{\rm CP}\mathbf{X}^{\rm mlp}_{\rm CP} + \mu_{\rm CP},
\end{equation}
where $\mu_{\rm CP}$ and $\sigma_{\rm CP}$ is the mean and variance of the channel matrix.
Then the tensors are rearranged to $\mathbf{X}^{O}_{\rm CP} \in \mathbb{R}^{T_2 \times M \times 2}$ to separate the real part and the imaginary part of the predicted channel.

\section{The Fine-tuning strategy of the Proposed Physical Layer Multi-task LLM}\label{sec-train}
Considering the huge parameter size of LLMs, directly fine-tuning all the parameters of the proposed network is impractical. To address this issue, this section focuses on the computationally efficient fine-tuning of the proposed PHY multi-task LLM.
First, we introduce LoRA~\cite{lora} for LLM fine-tuning, which inserts {\it low-rank adapters} into the LLM backbone to finetune the pre-trained LLM for PHY tasks.
Then, to mitigate the computational and memory demands of the proposed model, we employ a LoRA fine-tuning-aware {\it quantization method}~\cite{loftq} that performs LLM quantization while concurrently properly initializes the low-rank adapters for LoRA-based fine-tuning.
Moreover, the multi-task loss as well as training schedule are illustrated.

\subsection{LoRA for LLM Mainbody }\label{sec4-a}
LoRA~\cite{lora} is a parameter-efficient technique, which freezes the pre-trained model weights and injects trainable rank decomposition matrices into each layer. The use of LoRA mainly brings two advantages. Firstly, due to the immense parameter size of LLMs, full-parameter fine-tuning could be impractical.
LoRA circumvents this by only fine-tuning low-rank weight matrices, with a significantly reduced number of parameters. Secondly, LoRA prevents the problem of catastrophic forgetting of the original knowledge during fine-tuning. This is attributed to the fact that the newly learned knowledge has a lower rank than the original knowledge. As a result, LoRA facilitates the use of the universal modeling and generalization capability of pre-trained LLMs to achieve multiple tasks with a single model.

Specifically, we focus on the fine-tuning of the query matrices and value matrices in an attention block of the LLM backbone, denoted as  $\mathbf{W}_q$ and $\mathbf{W}_v$ respectively, as illustrated in Fig.~\ref{pic_lora}. 
LoRA updates two rank decomposition weight matrices $\mathbf{A}$ and $\mathbf{B}$ that are attached to a frozen pre-trained weight matrix $\mathbf{W}$ ($\mathbf{W}_q$ and $\mathbf{W}_v$). 
In this case, the original weight $\mathbf{W}$ is modified as
\begin{equation}
    \mathbf{W}  \leftarrow \mathbf{W}+\mathbf{A} \mathbf{B}^T,
\end{equation}
where $\mathbf{W} \in \mathbb{R}^{d_1 \times d_2}$, $\mathbf{A} \in \mathbb{R}^{d_1 \times r}$, $\mathbf{B} \in \mathbb{R}^{d_2 \times r}$, and $r \ll \min\{d_1, d_2\}$. Generally, we initialize the weights as follows:
\begin{equation}\label{eq-initial}
    \mathbf{A} \sim \mathcal{N}(0, \sigma^2), \quad \mathbf{B} = 0.
\end{equation}
As mentioned above, during the fine-tuning, we freeze $\mathbf{W}$ while update $\mathbf{A}$ and $\mathbf{B}$.
Since $r<< \min(d_1, d_2)$, the number of parameters for fine-tuning $\mathbf{A}$ and $\mathbf{B}$, i.e., $rd_1 + rd_2$, is significantly less than that of the full weight matrix, $d_1d_2$. The reduced parameter size thus makes the LoRA-based fine-tuning much efficient. 

\subsection{LoRA Fine-tuning-aware Quantization}
Despite the usage of LoRA, the extensive computational and memory demands of LLM-based models still pose significant challenges for both fine-tuning and inference, especially for resource-restricted equipment in wireless communications. 
To reduce the storage demands of pre-trained models, quantization acts as an essential compression strategy. Its key idea is to transform the original weights with a high precision into a finite set of discrete values.  For instance, converting model parameters from the original 16-bit floating-point format (FP16) to a 4-bit integer format leads to a 75\% decrease in storage overhead. 
However, it is worth noting that a quantized network parameters might degrade the performance of the aforementioned LoRA fine-tuning strategy. 
Particularly, consider a quantized weight matrix $\mathbf{Q}=q(\mathbf{W})$, where $q(\cdot)$ denotes the quantization operator. If the low-rank adapters $\mathbf{A}$ and $\mathbf{B}$ initialized by (29) and then attached to $\mathbf{Q}$, the initial weight matrix $\mathbf{Q}+\mathbf{A}\mathbf{B}^T$ naturally diverges from the original pre-trained parameters $\mathbf{W}$ because of the variations caused by the quantization process. Such unavoidable variations may negatively influence the initial setup of the LoRA fine-tuning procedure.

To solve the problem, we adopt a LoRA-fine-tuning-aware method to smoothly integrate quantization into the procedure of LoRA fine-tuning. The main idea is to approximate the original high-precision pre-trained weights by alternatively applying quantization for LLMs and proper low-rank initialization for LoRA fine-tuning. This initialization strategy effectively reduces the gap between the quantized and full-precision model, leading to a notable enhancement in fine-tuning performance.

\begin{figure}
	\centering 
	\includegraphics[width= \linewidth]{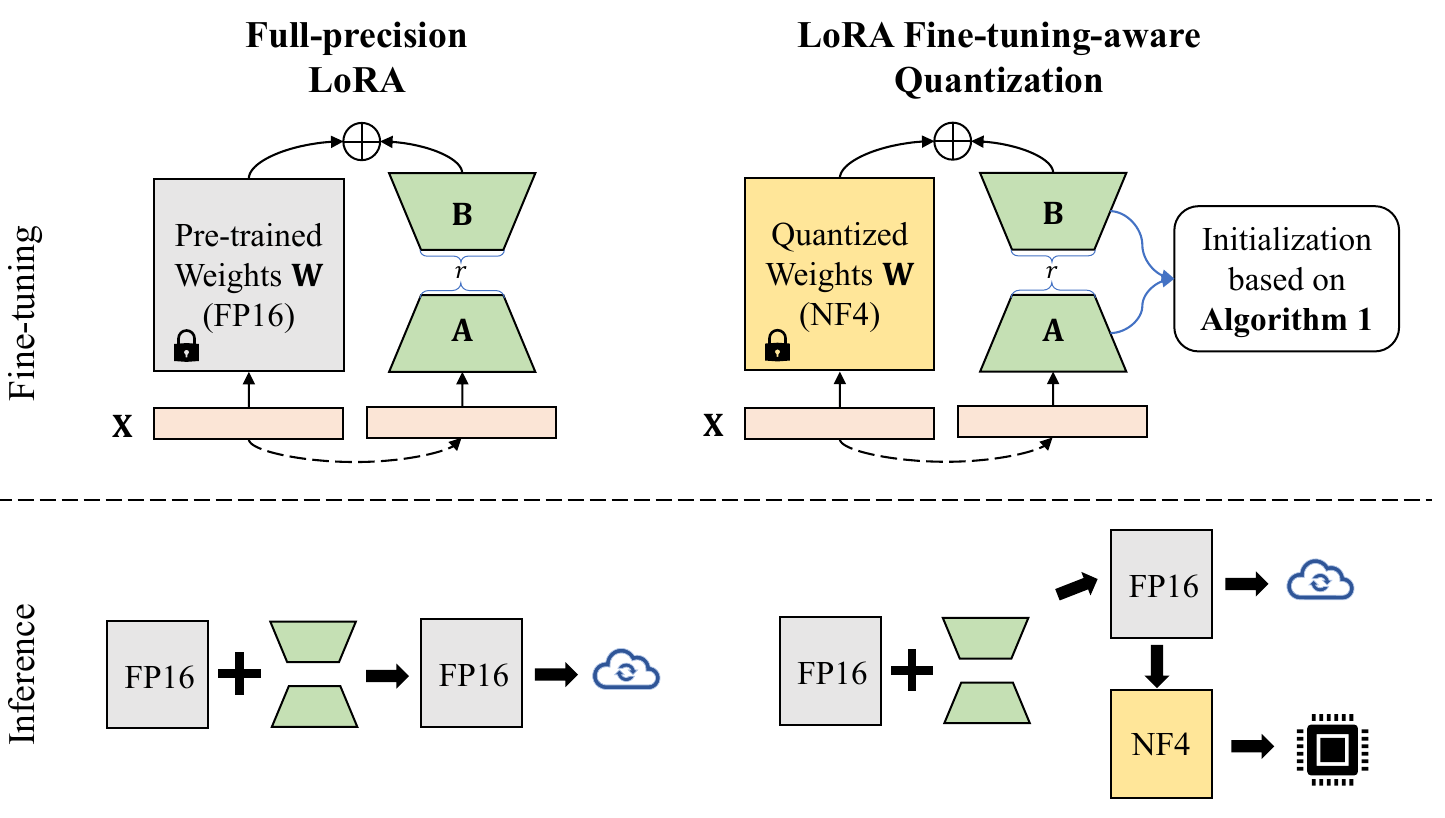}
        \caption{The comparison of LoRA fine-tuning-aware quantization with traditional full-precision LoRA.}
	\label{pic_quan}
\end{figure}

The key procedures of the LoRA fine-tuning-aware quantization method are presented in Fig.~\ref{pic_quan}.
During the initialization of LoRA fine-tuning, 
a quantized weight matrix $\mathbf{Q} \in \mathbb{R}^{d_1 \times d_2}$ with $N$-bit precision, along with the low-rank adapters $\mathbf{A} \in \mathbb{R}^{d_1 \times r}$, $\mathbf{B} \in \mathbb{R}^{d_2 \times r}$  are designed to closely approximate the original full-precision pre-trained parameter matrix $\mathbf{W} \in \mathbb{R}^{d_1 \times d_2}$. Mathematically, the model weight initialization problem can be formulated as:
\begin{equation}
    \textbf{P4: } \min_{\mathbf{Q},\mathbf{A},\mathbf{B}} \| \mathbf{W} - \mathbf{Q} - \mathbf{A}\mathbf{B}^T \|_F.
\end{equation}
This problem can be efficiently solved via alternatively conducting quantization and singular-value-decomposition. The step-by-step procedures provided in {\bf Algorithm 1} and the details are as follows. 

\textbf{Quantization Step:} In the $i$-th iteration, the quantization process is applied to the residual between the pre-trained parameter matrix $\mathbf{W}$ and the low-rank approximation $\mathbf{A}_{i-1}\mathbf{B}_{i-1}^T$, yielding the quantized weight matrix $\mathbf{Q}_i$
\begin{equation}
    \mathbf{Q}_i = q_N(\mathbf{W} - \mathbf{A}_{i-1}\mathbf{B}_{i-1}^T),
\end{equation}
where $q_N(\cdot): \mathbb{R} \mapsto \{0,1,\cdots, 2^N-1\}$ maps a high-precision weight matrix, e.g., matrix with 16-bit floating point number, to an $N$-bit quantized matrix. Typically, the quantization process can be expressed as
\begin{equation}
    \mathbf{Q} = {\rm round}((2^N-1)F(\mathbf{W})),
\end{equation}
where $F(\cdot): \mathbb{R} \mapsto [0,1]$ is a normalization function. In this work, we utilize the 4-bit NormalFloat Quantization (NF4) proposed in~\cite{qlora} to model the normalization function.  It assumes $\mathbf{W} \sim N(0,\sigma^2)$ and hence $F(\mathbf{W})=\Phi(\mathbf{W}/\sigma)$,  where $\Phi(\cdot)$ is the cumulative distribution function of the standard normal distribution. Besides, other quantization methods can also be involved such as the uniform quantization.

\textbf{SVD Step:}  After obtaining the $i$-th quantized weight $\mathbf{Q}_i$, SVD is applied to the residual of the quantization denoted by $\mathbf{R}_i =\mathbf{W}-\mathbf{Q}_i$ by
\begin{equation}
    \mathbf{R}_i = \sum_{j=1}^{d} \sigma_{i,j} {\bf u}_{i,j} {\bf v}_{i,j}^T,
\end{equation}
where $d = \min\{d_1, d_2\}, \quad \sigma_{i,1} \geq \sigma_{i,2} \geq \ldots \geq \sigma_{i,d}$ are the singular values of $\mathbf{R}_i$, ${\bf u}_{i,j}$, $ {\bf v}_{i,j}$ are the associated left and right singular vectors of $\mathbf{R}_i$. We then obtain a rank-$r$ approximation of $\mathbf{R}_i$ by $\mathbf{A}_{i-1}\mathbf{B}_{i-1}^T$, where
\begin{align}
    \mathbf{A}_{i} = [\sqrt{\sigma_{i,1}}{\bf u}_{i,1}, \ldots, \sqrt{\sigma_{i,r}}{\bf u}_{i,r}], \\
    \mathbf{B}_{i} = [\sqrt{\sigma_{i,1}}{\bf v}_{i,1}, \ldots, \sqrt{\sigma_{i,r}}{\bf v}_{i,r}].
\end{align}

After obtaining the initialization, LoRA fine-tuning can be performed as described in Section~\ref{sec3-c}. 

Moreover, during the inference, we merge the quantized backbone with the finetuned adapters to acquire the final output. If the deployed device is resource-limited, the merged model can be further quantized before deployment; otherwise, the full-precision model can be deployed directly.

\begin{algorithm}[htb]
	\caption{The initialization of LoRA fine-tuning-aware quantization.}
	\label{alg:ini}
	\renewcommand{\algorithmicrequire}{\textbf{Input:}}
	\renewcommand{\algorithmicensure}{\textbf{Output:}}
	\begin{algorithmic}[1] 
	\REQUIRE Full-precision pre-trained weight $\mathbf{W}$, LoRA rank $r$, $N$-bit quantization function $q_N(\cdot)$, iteration number $Iter$.\\
	\STATE   Initialize $\mathbf{A}_{0} \leftarrow   \mathbf{0},\mathbf{B}_{0} \leftarrow \mathbf{0}$;
        \STATE \textbf{for} $i = 1$ to $Iter$ \textbf{do}
        \STATE \quad  Obtain quantized weight $\mathbf{Q}_i \leftarrow q_N(\mathbf{W} - \mathbf{A}_{i-1}\mathbf{B}_{i-1}^T)$; 
        \STATE \quad  Obtain low-rank approximation $\mathbf{A}_{i},\mathbf{B}_{i} \leftarrow$ SVD$(\mathbf{W} -\mathbf{Q}_i)$ based on [36]-[37];
        \STATE \textbf{end for}
	\ENSURE $\mathbf{Q}_{Iter}$, $\mathbf{A}_{Iter}$, and $\mathbf{B}_{Iter}$.
	\end{algorithmic}
\end{algorithm}

\subsection{Multi-task Loss Function and Training Schedule}
During the training stage, we randomly choose a task and sample data from the corresponding dataset for each iteration. After determining the involved task and data, we select the proper prompt and activate the required modules of encoder and decoder. It is worth noting that the backbone of the pre-trained LLM is frozen, while the other parameters of the network, together with the LoRA adapters are trainable. Then the proposed network update the parameters of encoders, decoders and LoRA adapters using the corresponding loss. The detailed loss functions for each task are illustrated below. 

For \textbf{multi-user precoding}, the training is divided into two stages, namely supervised learning and unsupervised learning, respectively. The WMMSE algorithm is one of the most popular iterative algorithms to find the locally optimal solutions. Therefore, in the supervised learning stage, we can first generate the power allocation vectors $\underline{\mathbf{p}}$ and $\underline{\mathbf{\lambda}}$ using the WMMSE algorithm as the label. 
Then, the supervised learning will employ the MSE loss function to make the power allocation vectors, $\hat{\mathbf{p}}$ and $\hat{\mathbf{\lambda}}$, generated by LLM as close to $\underline{\mathbf{p}}$ and $\underline{\boldsymbol{\lambda}}$ as possible, i.e., 
\begin{equation}\label{eq:wmmmse}
    \text{Loss}_{Pre} = \frac{1}{2K} \left( \| \underline{\mathbf{p}} - \hat{\mathbf{p}} \|_2^2 + \| \underline{\boldsymbol{\lambda}} - \hat{\boldsymbol{\lambda}} \|_2^2 \right).
\end{equation}

Nevertheless, the WMMSE algorithm achieves only local optimality, making~\eqref{eq:wmmmse} insufficient for fully addressing the fundamental objective of problem \textbf{P3}. To enhance the overall rate performance, we implement additional network training using an unsupervised learning manner, where the loss function is directly derived from the original optimization objective:
\begin{equation}
     \text{Loss}_{Pre} = -\sum_{k=1}^{K} \log_2(1 + \gamma_k)
\end{equation}

In the \textbf{signal detection} task, 
the original transmitted data is rearranged as $\mathbf{X}_{\rm Det} \in \mathbb{R}^{K \times 2}$, which serves as ground truth of the network output.
Then we choose MSE loss for the training,
\begin{equation}
    \text{Loss}_{Det} = \frac{1}{2K}  \|\mathbf{X}_{\rm Det} - \mathbf{X}_{\rm Det}^{O} \|_F^2.
\end{equation}

For \textbf{channel prediction}, the ground truth of the predicted CSI is also available. We transform the complex CSI matrix to real ground truth $\mathbf{X}_{\rm CP} \in \mathbb{R}^{T_2 \times M \times 2}$, and MSE is adopted as the loss function to minimize the prediction error, i.e.,
\begin{equation}
        \text{Loss}_{CP} = \frac{1}{2MT_2}  \|\mathbf{X}_{\rm CP} - \mathbf{X}_{\rm CP}^{O} \|_F^2.
\end{equation}

\section{Simulation Results}\label{sec-sim}
In this section, extensive numerical simulations are presented to verify the effectiveness of the proposed method. Firstly, we elaborate on the simulation setup. Then the performance of three selected tasks, i.e. channel prediction, multi-user precoding, and signal detection, is evaluated respectively. Besides,  the performance of LoRA fine-tuning-aware quantization is compared with the full-precision model. Finally, we analyze the impact of designed multi-task instruction and different LLM backbones, respectively.

\subsection{Simulation Setup and Training Details}
For the experimental setup, we utilize the QuaDRiGa channel generator~\cite{QuaDRiGa}, implementing the 3GPP Urban Macro (UMa) propagation model~\cite{3gpp} under non-line-of-sight (NLOS) conditions. The channel consists of 21 scattering clusters, each containing 20 propagation paths.
We consider a multi-user MISO-OFDM system, where a BS simultaneously serves $K=4 \sim 8$ moving users. BS employs a UPA comprising $N_h = 16$ elements in horizontal and $N_v = 8$ in vertical, while users are configured with single-antenna receivers. The antenna spacing is maintained at half of the wavelength at the center frequency. 
The users are uniformly distributed within angle range $[-\pi/2,\pi/2]$, and distance range $[\rho_{\rm min}, \rho_{\rm max}] = [20~{\rm m}, 100~{\rm m}]$.
We suppose a time-division duplex (TDD) system, where the center frequency of channel is set as $2.4$ GHz. The bandwidth of the channel is 8.64 MHz, comprising of $M=48$ subcarriers, i.e., the frequency interval of  subcarriers is $180$ kHz. The dataset is partitioned into training and testing subsets, containing 50,000 and 10,000 samples per task respectively. The model undergoes training for 200 epochs over the dataset.

For channel prediction problem, We predict future CSI of $T_2 = 4$ time slots based on historical CSI of $T_1 = 16$ time slots. We suppose that the users are initialized with random positions and follow linear movement patterns. The velocity distribution for mobile users spans uniformly from $10$ km/h to $100$ km/h. To enhance the robustness against noise, the SNR is uniformly sampled between 5 dB and 20 dB during the training stage, to account for both the low-noise and high-noise scenarios.
For uplink signal detection, we suppose the transmitted data is generated from 16-QAM modulation symbol.
Similarly, the received signal is corrupted by noise with SNR uniformly distributed between 5 and 20 dB, during the fine-tuning.

\subsection{Performance Evaluation for Channel Prediction}
\subsubsection{Baselines and Performance Metric}
To evaluate the performance, we compare the proposed multi-task LLM with the following benchmarks.
\begin{itemize}
    \item  Transformer: \cite{Informer} introduces a parallelized channel prediction framework based on transformer to predict future CSI in parallel, and thus avoid error propagation problems. 
    \item RNN: A Recurrent Neural Network (RNN)~\cite{RNN} is a traditional type of neural network used for processing sequences and is commonly utilized in channel prediction tasks. In the experiments, we configure the RNN with four layers.
    \item  LSTM: 
    The long short-term memory network (LSTM)~\cite{lstm} incorporates specialized memory units and gating mechanisms to effectively capture long-range temporal dependencies. Our implementation utilizes a four-layer LSTM structure for predictive modeling.
    \item  GRU: As an enhanced version of LSTM, the gated recurrent unit (GRU)~\cite{GRU} introduces simplified gating operations to mitigate gradient-related challenges during training. Similarly, the GRU-based model consists of 4 layers.
    \item LLM4CP:
    LLM4CP ~\cite{LLM4CP} represents a pioneering effort in leveraging pre-trained language model, i.e. GPT-2, for channel prediction task, demonstrating superior prediction accuracy and enhanced generalization.
    \item Single-task LLM: We also train the proposed network on a single task to better compare the performance with the proposed multi-task LLM.
\end{itemize}

In channel prediction evaluation, the NMSE serves as a crucial indicator for assessing prediction precision, making it an essential measurement criterion in our experimental analysis.

\subsubsection{Performance Analysis}
\begin{figure}
	\centering 
	\includegraphics[width= \linewidth]{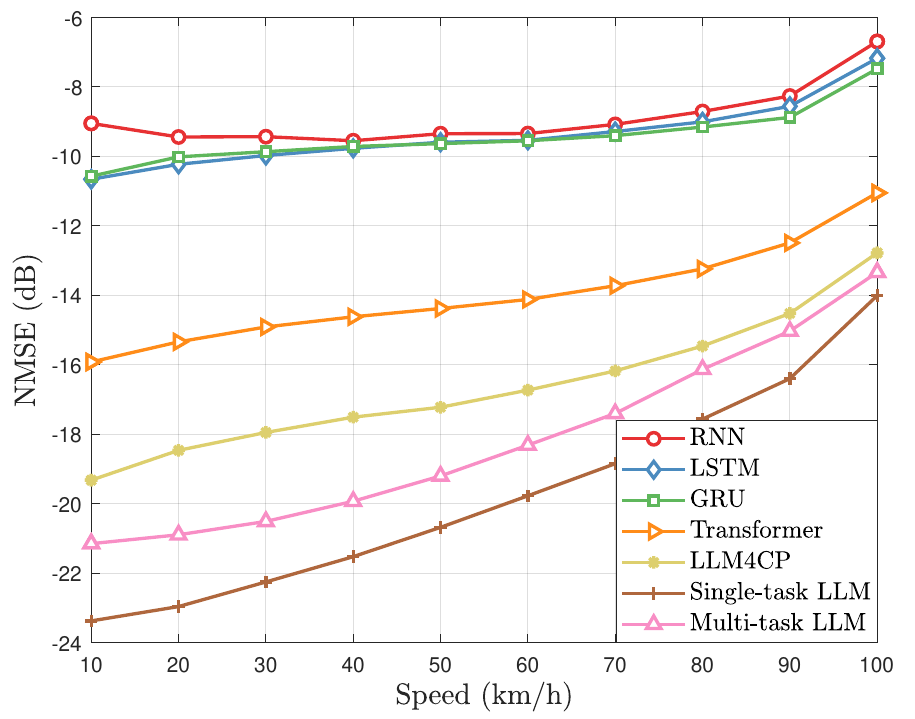}
        \caption{The NMSE performance of proposed method and other baselines versus different user velocities.}
	\label{pic_cp}
\end{figure}

The evaluation dataset for channel prediction comprises 10 distinct velocities spanning from $10$ km/h to $100$ km/h, with each velocity containing $1000$ data samples. As illustrated in Fig.~\ref{pic_cp}, the NMSE performance of our proposed multi-task LLM framework is compared against various baseline methods across different user velocities. The historical CSI data is added by Gaussian whiten noise with ${\rm SNR} = 20$ dB.
Experimental results demonstrate a consistent degradation in NMSE performance across all methods as user mobility increases. This phenomenon can be attributed to the accelerated channel variation and reduced coherence time associated with higher velocities, which consequently amplifies the complexity of accurate channel estimation.
Fig.~\ref{pic_cp} reveals that attention-based methods, achieve relatively high performance than traditional AI methods, validating the potential of attention-based methods in channel prediction task. Equipped with excellent modeling capability of LLM, the proposed model finetuned on channel prediction task only, consistently outperforms other baselines among testing velocities. With elaborately designed multi-task instruction, the proposed LLM-enabled multi-task model obtains comparable channel prediction accuracy with the single-task one, which verified the effectiveness of the proposed multi-task LLM. 

\begin{figure}
	\centering 
	\includegraphics[width= \linewidth]{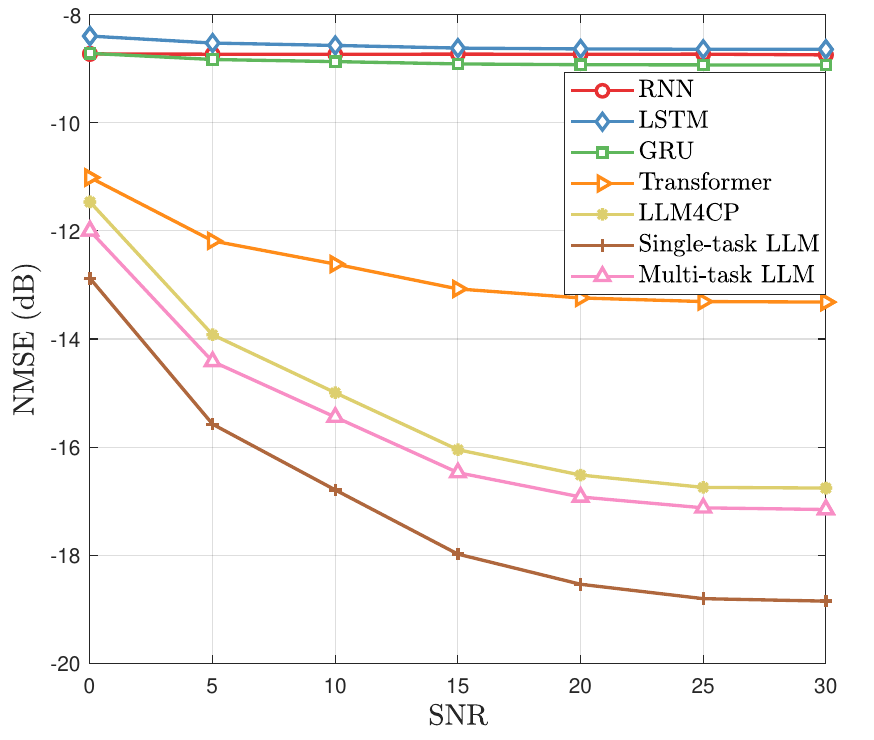}
        \caption{The NMSE performance of proposed method and other baselines versus different SNR.}
	\label{pic_cp_snr}
\end{figure}

In Fig.~\ref{pic_cp_snr}, the robustness against noise of the proposed method is evaluated, where the SNR of noise in historical CSI is growing from 0 dB to 25 dB. The NMSE performance has been averaged over all test speeds. It can be observed that for all schemes, increased SNR conditions lead to improved NMSE performance in prediction accuracy. Thanks to the generalization ability of LLMs, the proposed method exhibits high robustness performance against CSI noise. It achieves the lowest NMSE performance in the entire SNR regime.

\subsection{Performance Evaluation for Multi-user Precoding}
\subsubsection{Baselines and Performance Metric}
For multi-user precoding, the following methods are selected as baselines, including traditional methods and deep learning-based methods.
\begin{itemize}
    \item ZF: Eigen-based zero-forcing (ZF) algorithm~\cite{prezf} is a computationally efficient approach, which derives the precoding matrix through the Moore-Penrose pseudo-inverse operation applied to the channel matrix.
    \item WMMSE: As mentioned above, the WMMSE algorithm~\cite{WMMSE} is one of the most popular iterative algorithms. The method can achieve satisfactory sum rate performance while suffering from high computational complexity. 
    The iterative number is set as $20$.
    \item CNN: In~\cite{precnn}, the authors propose a CNN-based framework for the optimization of downlink beamforming.
    \item Single-task LLM: Similarly, we train the proposed network on multi-user precoding only.
\end{itemize}

The sum rate of users, which is the objective of multi-user precoding, is utilized as performance metric to evaluate the multi-user precoding task.

\subsubsection{Performance Analysis}
\begin{figure}
	\centering 
	\includegraphics[width= \linewidth]{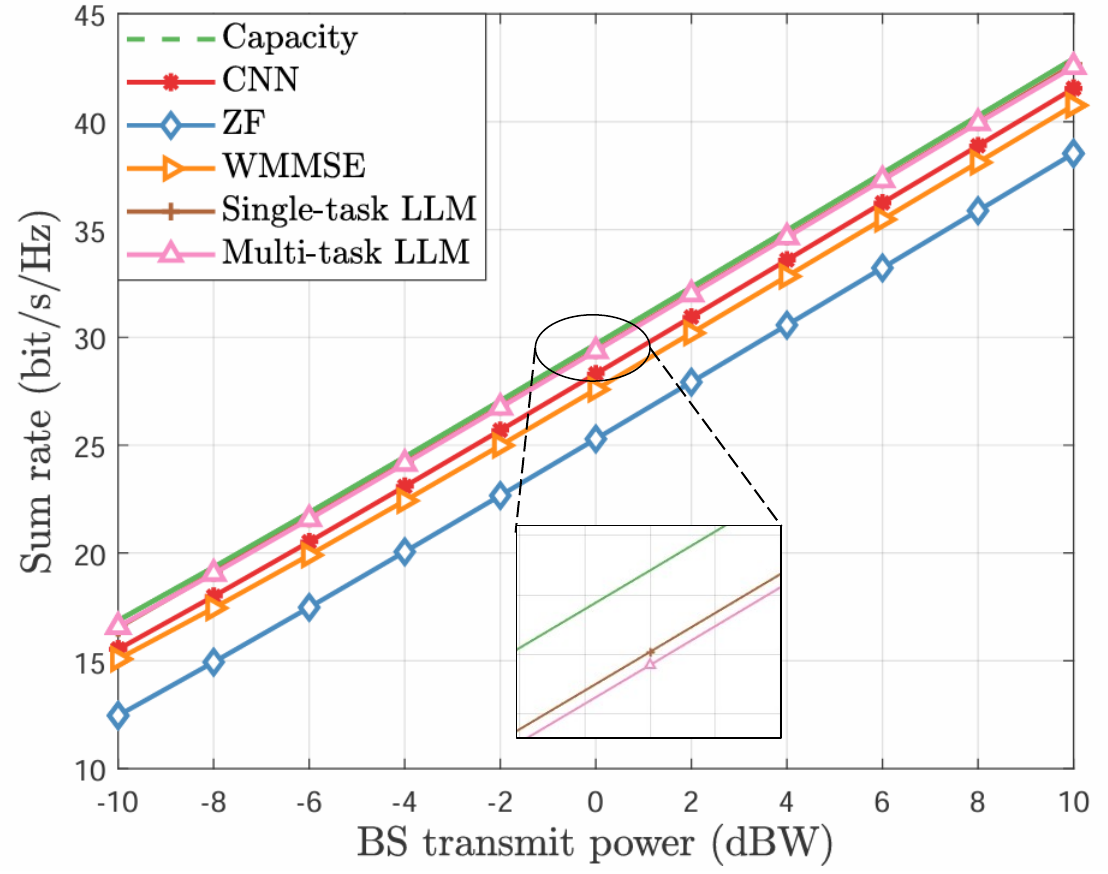}
        \caption{The sum rate performance of proposed method and other baselines versus different BS transmit powers.}
	\label{pic_pre}
\end{figure}

The sum rate performance against different BS transmit power is plotted in Fig.~\ref{pic_pre}.
The noise power is set as $\sigma^2 = -10$ dBW, the maximal transmit power of BS $P_{\rm max}$ increases from $-10$ to $10$ dBW. Besides, the user number is set as 4.
As illustrated in Fig.~\ref{pic_pre}, with the increase of the BS transmit power, the sum rate of all methods increase accordingly.
As depicted in Fig.~\ref{pic_pre}, the ZF-based method, though with low complexity, achieves unsatisfactory performance. 
The iterative algorithm WMMSE improves the sum rate, while it is still possible to fall in local optimal solutions, inducing an obvious gap from the capacity.
With the powerful network and carefully designed training strategy, the deep-learning-based methods, including the CNN-based method and LLM-based method, are promising in conquering the problem and further improving the performance. 
Specifically, the proposed model, both trained on a single task and trained on multiple tasks, achieves near-optimal performance for all transmit power and outperforms other benchmark schemes for the entire transmit power range. Owing to the increasing size of the network, the LLM-based method exhibits superior optimization and generalization capabilities and outperforms the CNN-based method in terms of performance.

\begin{figure}
	\centering 
	\includegraphics[width= \linewidth]{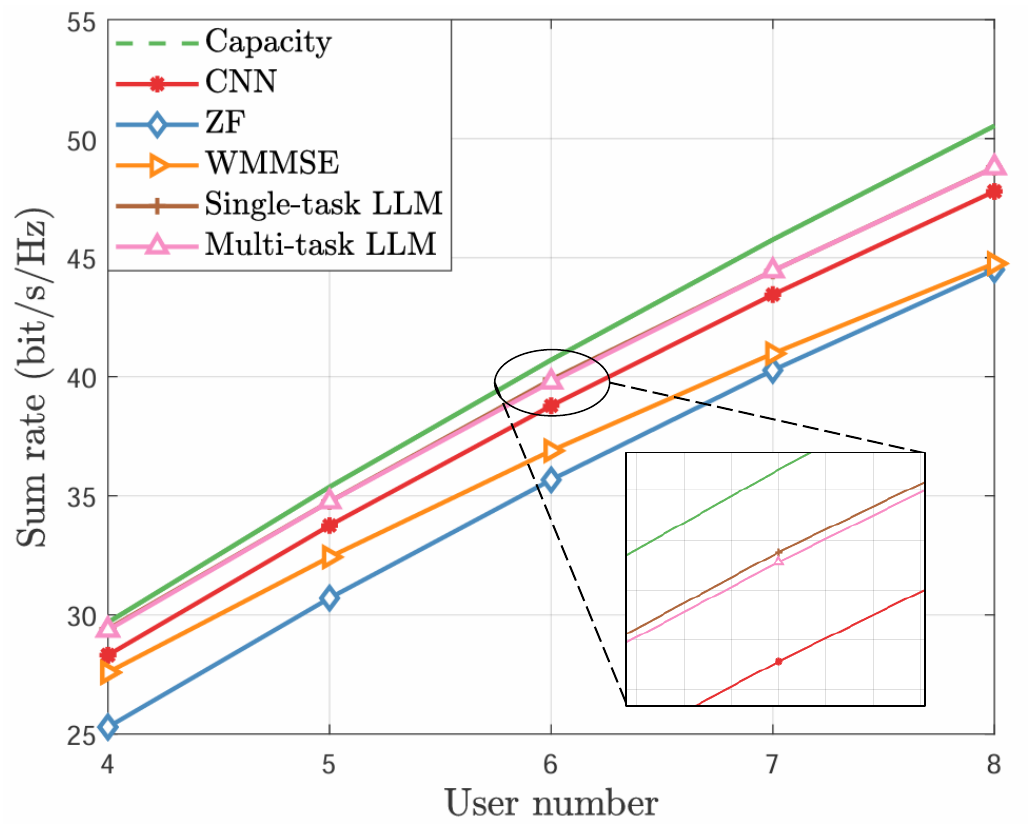}
        \caption{The sum rate performance of proposed method and other baselines versus different user numbers.}
	\label{pic_pre_user}
\end{figure}

In Fig.~\ref{pic_pre_user}, we illustrate the sum rate performance against the user number, which ranges from 4 to 8. The transmit power and noise power is set as 0 dBW and -10 dBW, respectively. As the user number increases, the spectrum efficiency increases with further exploitation of multiplexing gain. 
We observe from Fig.~\ref{pic_pre_user} that the proposed multi-task LLM-based method enjoys a higher sum rate performance, compared to existing methods for different number of users. This verifies the effectiveness and scalability of the proposed method.
It is worth noting that LLMs inherently possess the ability to process variable-length sequences, thus our proposed scheme can be directly applied to different number of users without requiring any modifications. In contrast, for traditional AI methods, to accommodate varying numbers of users, input data under different user numbers must be zero-padded to match the shape of the maximum user number.

\subsection{Performance Evaluation for Signal Detection}
\subsubsection{Baselines and Performance Metric}
To validate the effectiveness of the proposed method, several methods are implemented as baselines.
\begin{itemize}
    \item LMMSE: Linear minimum mean-squared error (LMMSE) detector is a classical method for achieving signal detection with low complexity.
    \item DNN: In~\cite{DECDNN}, a data-driven model which inputs the channel as well as the received data and outputs the original data through a deep learning network.
    \item DetNet: The detection network (DetNet) in~\cite{detnet} unfolds the iterations of a projected gradient descent algorithm to recover the data.
    \item OAMP-Net: A famous model-driven deep learning network proposed in~\cite{OAMP}, which incorporates deep learning into the orthogonal AMP (OAMP) algorithm for accurate signal detection.
    \item Single-task LLM: The proposed network is finetuned only on the signal detection task.
\end{itemize}

In this work, we utilize NMSE and symbol error ratio (SER) as metrics to evaluate the performance of signal detection. 
NMSE loss is the direct performance metric to present the accuracy of data recovery. 
It should be noted that the reason for employing the data recovery accuracy rather than only choosing the bit error ratio (BER) or SER to optimize and evaluate the model is described below. In the practical communication link, the objective of signal detection is to recover the data as accurately as possible, which can be utilized to compute the log-likelihood ratio (LLR) for demodulation and channel decoding to enable reliable communications.
Besides, obtaining the recovered signal, we also demodulate the transmitted symbol by minimum euclidean distance decision, and then SER is employed to evaluate the performance in this case.

\subsubsection{Performance Analysis}
\begin{figure}
	\centering 
	\includegraphics[width= \linewidth]{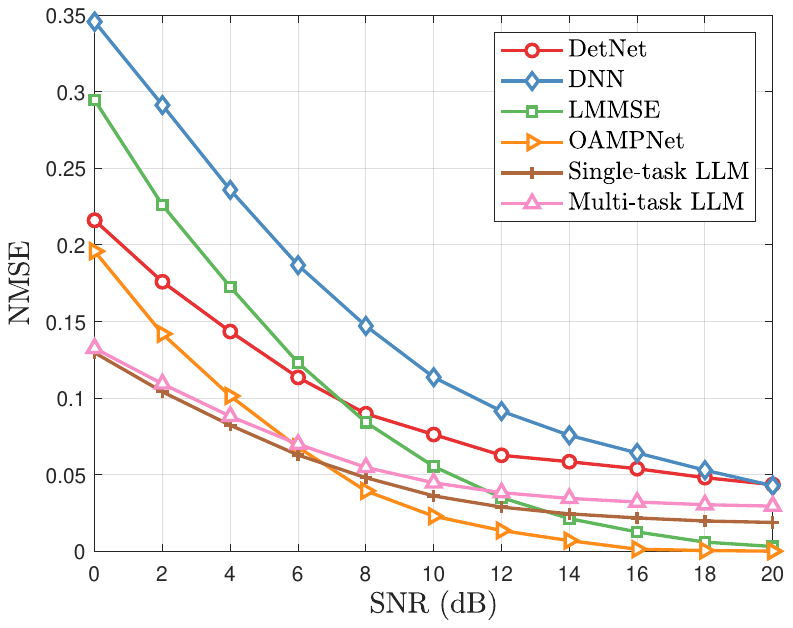}
        \caption{The NMSE performance of the proposed method and other baselines versus different SNRs.}
	\label{pic_decnmse}
\end{figure}

In Fig.~\ref{pic_decnmse}, the NMSE performance of the proposed multi-task LLM on signal detection under different SNRs is presented. We also compare the performance with several baselines. 
As illustrated in Fig.~\ref{pic_decnmse}, the DNN-based method presents poor performance since it directly inputs the data to the black-box-based network, while ignores the domain knowledge and structure of the problem and data. Besides, despite the low computational complexity, the LMMSE-based method also shows relatively high NMSE performance. The model-based methods, including DetNet~\cite{detnet} and OAMPNet~\cite{OAMP}, significantly improve the data recover accuracy; the OAMPNet~\cite{OAMP} achieves higher performance than the LLM-based method in high SNR regime with better utilization of the statistical information of noise.
Moreover, the proposed multi-task LLM can accurately recover the transmitted data, especially in the low SNR regime. It can be observed that the performance of multi-task LLM is comparable to that of single-task LLM, suggesting the potential of multi-task LLM networks in wireless communications.

\begin{figure}
	\centering 
	\includegraphics[width= \linewidth]{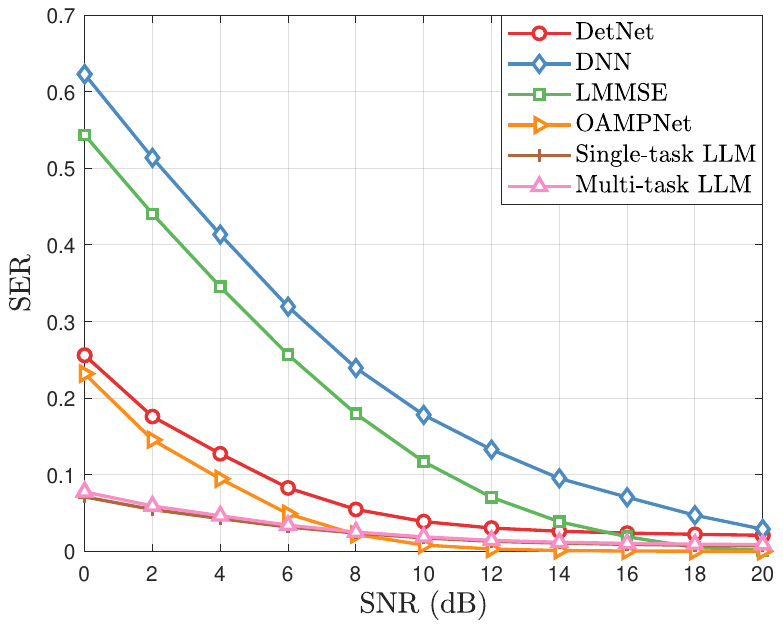}
        \caption{The SER performance of the proposed method and other baselines versus different SNRs.}
	\label{pic_ser}
\end{figure}

The SER performance against SNR is depicted in Fig.~\ref{pic_ser}. When the SNR is lower than 8 dB, the proposed multi-task LLM network outperforms other baselines, which indicates that the generalization capability of LLMs endows the proposed method with high robustness against noise. As the SNR increases, the OAMPNet~\cite{OAMP} achieves the best performance, since it fully utilizes the statistical information of noise. Based on this observation, we provide two comments as follows. First, for the scenarios where the power of noise $\sigma^2$ can be obtained, future works can consider effectively incorporating statistical information of channel and noise as prompt to further improve the performance of LLM-based method, especially in high SNR regimes. Secondly, there are still many cases where the statistical information of noise may not be obtained in practical system. In these cases, OAMPNet may fail to achieve satisfactory performances.

\subsection{Performance evaluation for LoRA-fine-tuning-aware quantization} \label{sec-loraquan}
For fair and convenient comparison, in the above subsections, we mainly utilize full-precision model for performance evaluation. Then in this subsection, we focus on the impact of LoRA-fine-tuning-aware quantization.

\begin{table*}[htb]
\centering
\caption{Comparison of the model with LoRA-fine-tuning-aware quantization with other baselines.}
\begin{tabular}{c|ccc}
\toprule
\textbf{Quantization method}                     & \textbf{NMSE for CP} & \textbf{NMSE for DET} & \textbf{Sum rate for PRE} \\ \hline
Full-precision                      & -16.37 dB             & 0.0322                & 29.3658 bit/s/Hz                  \\ \hline
LoRA-fine-tuning-aware quantization & -16.25 dB             & 0.0363                & 29.3632 bit/s/Hz                   \\ \hline
Traditional quantization            & -15.68 dB             & 0.0528                & 28.6032 bit/s/Hz                  \\ \bottomrule
\end{tabular}
\vspace{0.5em}

\footnotesize 
    $\bullet$  ``CP" denotes channel prediction, ``DET" denotes signal detection, and ``PRE" denotes multi-user precoding.
\end{table*}\label{tab-quan}

\begin{table*}[htb]
\centering
\caption{Comparison of the model with different LLM backbone.}
\begin{tabular}{c|cccc}
\toprule
\textbf{LLM backbone}  & \textbf{NMSE for CP} & \textbf{NMSE for DET} & \textbf{Sum rate for PRE} & \textbf{Network parameters}  \\ \hline
LLAMA2-7B  & -16.37 dB             & 0.0322                & 29.3658 bit/s/Hz  & 23.9713 M    \\ \hline
GPT2 & -15.39 dB             & 0.0573               & 29.0365 bit/s/Hz  & 17.3338 M    \\ \hline
-  & -13.49 dB             & 0.1908               & 27.2981 bit/s/Hz   & 19.777 M    \\ \bottomrule
\end{tabular}
\vspace{0.5em}

\footnotesize 
    $\bullet$  ``CP" denotes channel prediction, ``DET" denotes signal detection, and ``PRE" denotes multi-user precoding.
\end{table*}\label{tab-backbone}

The performance comparison of the proposed full-precision model, the model with LoRA-fine-tuning-aware quantization, and the model with traditional quantization are presented in Table I.
For the full-precision model, the model parameters are stored and computed in a 16-bit floating-point format.
On the contrary, the model with LoRA-fine-tuning-aware quantization transforms the LLM backbone into a 4-bit integer format, which indicates a 75\% storage reduction during the fine-tuning. Besides, the iterative number for initialization is set as $Iter = 5$. It is noted here that during LoRA fine-tuning,  the quantized weight is temporarily dequantized to the simulated high-precision weight to facilitate accurate computation. 
Moreover,  the model with traditional quantization serves as a baseline, which quantifies the LLM backbone weight matrices, initials the adapters as~\eqref{eq-initial}, and conducts LoRA fine-tuning afterward.

As shown in Table I, the model with LoRA-fine-tuning-aware quantization achieves comparable performance with the proposed full-precision model. 
For the channel prediction task, the NMSE is averaged for different speeds and SNRs; for the signal detection task, the performance is averaged for different SNRs; for multi-user precoding task, we set user number as 4, the power of noise and transmit power is set as -10 dBW and 0 dBW, respectively.
Besides, it significantly outperforms the model with traditional quantization for all the tasks selected. We can conclude from the results that the introduced LoRA-fine-tuning-aware quantization successfully approximates high-precision weight by the quantized weight and low-rank adapters, and thus the performance degradation resulting from quantization is negligible.


\subsection{Impact of multi-task instruction}
In this subsection, we analyze the influence of the proposed multi-task instruction module in Section~\ref{sec3-a} on the performance and convergence rate of neural network training.
It is worth noting that, in order to distinguish different tasks, utilizing multi-task instruction as prompts is indispensable for the proposed multi-task PHY network to understand different task requirements.
For single-task LLM, although it is possible to neglect the prompt, introducing the designed instruction with wireless data can involve domain knowledge to facilitate task-specific adaptation of LLMs.

\begin{figure}
	\centering 
	\includegraphics[width= \linewidth]{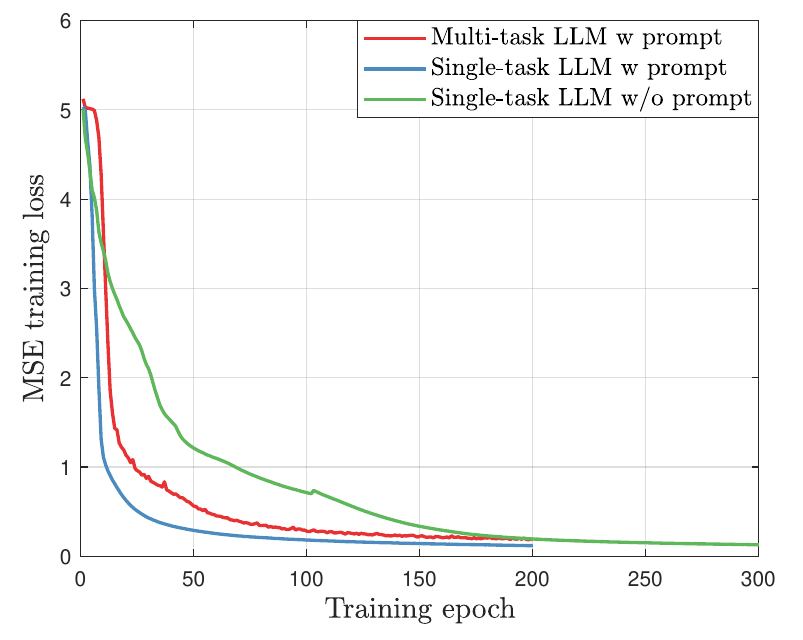}
        \caption{Training losses of different methods against epoch.}
	\label{pic_trainingloss}
\end{figure}

In Fig.~\ref{pic_trainingloss}, we take the signal detection task for instance to illustrate the training loss in MSE against the training epoch. In this figure, we depict the training loss of the proposed model trained for multiple tasks with prompts, the proposed model trained for a single task with prompts, and the proposed model trained for a single task without prompts. We can observe that the overall trends of all losses are declining with the increase of training epoch. The training loss of single-task LLM with prompts converges the fastest, and 200 epochs of training are enough to achieve satisfactory performance. When the training epoch reaches about 200, the loss function of multi-task LLM also converges. Nevertheless, the proposed model trained for a single task without prompts requires 300 epochs of training to obtain comparable performance with the model with prompts. Therefore, it is indicated that the incorporation of designed multi-task instruction as prompts significantly accelerates the network fine-tuning and improves the LLM’s adaptability to downstream tasks.

\subsection{Impact of LLM backbone}
Finally, in this subsection, the influence of LLM backbone on the performance is evaluated. In this work, we employ LLAMA2-7B as the backbone. To analyze the impact, we compare the selected backbone with the following benchmarks. Firstly, we replace the LLAMA2-7B with a smaller pre-trained model GPT2~\cite{gpt} while other modules remain the same.
Secondly, the LLM backbone is directly removed, which means that the output of the encoders is straightly fed into the designed decoders.

We summarize the simulation results in Table II.
Similarly, as stated in Section~\ref{sec-loraquan}, the NMSE is averaged within different SNRs and speeds for channel prediction and is averaged within different SNRs for signal detection. Besides, user number is set as 4, the power of noise and transmit power is set as -10 dBW and 0 dBW for multi-user precoding.
It can be observed from Table II that the performance achieved in different tasks improves with the introduction of the LLM backbone. For instance, the NMSE performance for channel prediction of the proposed method with LLAMA2 backbone is -16.37 dB, while the performance drops 3 dB when removing the backbone.
And the performance drops sharply especially for signal detection task.
Furthermore, due to the increasing size of the LLM backbone, the model with LLAMA2-7B backbone outperforms the model with GPT2 backbone, although GPT2 can still achieve the multi-task PHY network. Therefore, we can adopt a proper LLM backbone to balance the computational costs and performance. 

\section{Conclusions} \label{sec-con}
In this paper, we propose a LLM-enabled multi-task PHY network to unify multiple tasks with a single LLM. Multi-task instruction module, input encoders, as well as output decoders, are elaborately designed to distinguish multiple tasks and adapted the features of different formats of wireless data for the feature of LLM. 
Moreover, to reduce the memory requirements of the proposed model, a LoRA fine-tuning-aware quantization method is introduced.
Simulation results have verified the effectiveness of the proposed method. 
The proposed LLM framework is promising to perform different tasks using a single model, significantly saving the redundancy and costs of the practical deployment of LLM. 
It makes an initial attempt to provide a more adaptable, comprehensive, and intelligent PHY network with the aid of LLMs.
Future works can be focused on incorporating more tasks into the network. Besides, the LLM pruning and knowledge distillation techniques can also be employed to reduce the computational and storage overhead of LLM-based method.

\footnotesize

\bibliographystyle{IEEEtran}
\bibliography{multitask, IEEEabrv}

\normalsize



\end{document}